\documentclass{aastex62}

\usepackage{graphicx}
\usepackage{latexsym}
\usepackage{amsfonts,amsmath,amssymb}
\usepackage{epstopdf}
\usepackage{natbib}
\usepackage{mathrsfs}
\usepackage{algorithm}
\usepackage[utf8]{inputenc}
\usepackage{hyperref}
\usepackage[noend]{algpseudocode}

\def\apjref#1;#2;#3;#4 {\par\pni\ #1,  #2, {\bf #3}, #4. \par}

\shortauthors{Gaches, Offner \& Bisbas}
\newcommand{\COL}		{N({\rm H_2})}
\def\kms{\ifmmode {\rm km s}^{-1} \else km s$^{-1}$\fi}
\def\XCO{\ifmmode {X_{\rm CO}} \else X$_{\rm CO}$ \fi}
\def\XCI{\ifmmode {X_{\rm CI}} \else X$_{\rm CI}$ \fi}

\defcitealias{gaches2019}{Paper~I}

\begin{document}
\title{The Astrochemical Impact of Cosmic Rays in Protoclusters II: CI-to-H$_2$ and CO-to-H$_2$ Conversion Factors}

\author[0000-0003-4224-6829]{Brandt A.L. Gaches}
\affiliation{Department of Astronomy, University of Massachusetts - Amherst, Amherst, MA 01003, USA}
\affiliation{I.~Physikalisches Institut, Universit\"at zu K\"oln, Z\"ulpicher Stra{\ss}e 77, Germany}
\email{gaches@ph1.uni-koeln.de}

\author[0000-0003-1252-9916]{Stella S.R. Offner}
\affiliation{Department of Astronomy, The University of Texas at Austin, Austin, TX 78712, USA}
\email{soffner@astro.as.texas.edu}

\author[0000-0003-2733-4580]{Thomas G. Bisbas}
\affiliation{Department of Physics, Aristotle University of Thessaloniki, GR-54124 Thessaloniki, Greece}
\affiliation{I.~Physikalisches Institut, Universit\"at zu K\"oln, Z\"ulpicher Stra{\ss}e 77, Germany}
\affiliation{National Observatory of Athens, Institute for Astronomy, Astrophysics, Space Applications and Remote Sensing, Penteli, 15236, Athens, Greece}
\email{tbisbas@auth.gr; bisbas@ph1.uni-koeln.de}

\correspondingauthor{Brandt Gaches}

\begin{abstract}
We utilize a modified astrochemistry code which includes cosmic ray attenuation in-situ to quantify the impact of different cosmic-ray models on the CO-to-H$_2$ and CI-to-H$_2$ conversion factors, \XCO and \XCI, respectively. We consider the impact of cosmic rays accelerated by accretion shocks, and show that clouds with star formation efficiencies greater than 2\% have $\XCO = (2.5 \pm 1)\times10^{20}$ cm$^{-2}$(K km s$^{-1}$)$^{-1}$, consistent with Milky Way observations. We find that changing the cosmic ray ionization rate from external sources from the canonical $\zeta \approx 10^{-17}$ to $\zeta \approx 10^{-16}$ s$^{-1}$, which better represents observations in diffuse gas, reduces \XCO by 0.2 dex for clusters with surface densities below 3 g cm$^{-2}$. We show that embedded sources regulate \XCO and decrease its variance across a wide range of surface densities and star formation efficiencies. Our models reproduce the trends of a decreased \XCO in extreme cosmic ray environments. \XCI has been proposed as an alternative to \XCO due to its brightness at high redshifts. The inclusion of internal cosmic ray sources leads to 1.2 dex dispersion in \XCI ranging from $2\times10^{20} < \XCI < 4\times10^{21}$ cm$^{-2}$ (K km s$^{-1}$)$^{-1}$. We show that \XCI is highly sensitive to the underlying cosmic ray model.
\end{abstract}

\section{Introduction}
Studying the properties of molecular clouds is crucial to understand star formation \citep{kennicutt2012}. The dominant constituent of molecular clouds is molecular hydrogen, (H$_2$), which is a perfectly symmetric molecule, rendering it largely invisible at the typical temperatures of molecular clouds. While observable in ultraviolet absorption against background sources, it can only be detected via emission in environments where the gas is excited to temperatures above a few hundred Kelvin. The second dominant species is neutral helium which remains inert in molecular clouds. Therefore, observational studies of molecular clouds largely rely on tracer species, namely emission from dust and molecules. The most important of these tracers is carbon monoxide (CO) \citep{bolatto2013}. CO has a relatively high abundance, canonically [CO/H$_2$] $\approx$ 10$^{-4}$ \citep{hollenbach1999}, making it the most abundant molecule after H$_2$ . The small dipole moment allows its rotational transitions to be easily excited at the cold temperatures of molecular clouds. A crucial CO observable is the J = (1-0) rotational transition at a rest frequency of 115.27 GHz.

It is common for the emission of the lowest rotational transition of CO to be used to measure the total molecular gas \citep{bolatto2013}. This is encoded in the CO-to-H$_2$  conversion factor, $\XCO$, and the related quantity $\alpha_{\rm CO}$. $\XCO$ is defined as: 
\begin{equation}
\label{eq:xco}
\XCO = \frac{\COL}{W_{\rm CO}(J = 1-0)},
\end{equation}
where $\COL$ is the H$_2$ column density in cm$^{-2}$ and $W_{\rm CO}(J = 1-0)$ is the CO flux in K km s$^{-1}$. The fiducial Milky Way (MW) value is $X_{\rm CO,MW} = 2 \times 10^{20}$ cm$^{-2}$ (K km s$^{-1}$)$^{-1}$ \citep{bolatto2013}. This conversion factor has been used to estimate gas mass in local, resolved studies of MCs and the molecular gas mass in high redshift galaxies \citep[e.g. the COLDz survey][]{riechers2019}. A significant number of studies, both observational and theoretical, have been devoted to measuring, modelling or applying \XCO. Prior work shows it varies with density, metallicity \citep{bell2006, shetty2011, lagos2012, narayanan2013, glover2016}, cosmic ray (CR) ionization rate (CRIR) \citep{bell2006, wolfire2010, bisbas2015, clark2015, glover2016, remy2017, papadopoulos2018} and the radiation field \citep{bell2006, wolfire2010, shetty2011, lagos2012, narayanan2013, clark2015, glover2016, gaches2018a, gong2018}. Previously, \cite{gaches2018a} found that far ultraviolet radiation feedback from forming stars can reproduce the higher \XCO values measured towards diffuse star-forming clouds in the outer galaxy.

Traditional one-dimensional photo-dissociation region (PDR) models have long predicted that neutral carbon will exist only in a thin transitional layer between ionized carbon and CO \citep{hollenbach1999}. However, observations show that forbidden line emission from neutral carbon covers similar spatial extents as CO \citep[e.g.][]{ikeda1999, kulesa2005, lo2014}. It is posited that forbidden line emission from neutral carbon is a good tracer of the gas mass \citep{papadopoulos2004, offner2014, glover2015, glover2016}. Synthetic observations of hydrodynamic simulations show that \XCI has a smaller dispersion than \XCO within a molecular cloud and is a better tracer in low metallicity gas which tends to become CO-dark \citep{offner2014, glover2015, glover2016}. Observational studies using \XCI as a tracer of gas mass performs as well as \XCO \citep{lo2014}. \XCI is defined analogously to \XCO:
\begin{equation}
\label{eq:xci}
{\rm X_{\rm CI} = \frac{N(H_2)}{W(CI)_{609 \, \mu m}}} ~~ {\rm cm^{-2}\, (K \, km \, s^{-1})^{-1}}
\end{equation}
where $W({\rm CI})_{\rm 609 \, \mu m}$ is the integrated flux of the $^3P_1\rightarrow ^3\!P_0$ transition at 609~$\mu$m.

\cite{gaches2019} (hereafter Paper I) presented a modified astrochemical code which includes CR attenuation in-situ. \citetalias{gaches2019} included CRs accelerated by accreting, embedded protostars and CR attenuation in one-dimensional astrochemical models of molecular clouds. We used the code to study the impact of changing the CR spectrum due to differing galactic environments and the effects of embedded CR sources for a subset of species including CO, HCO$^+$ and N$_2$H$^+$ and tested various prescriptions for constraining the CRIR. We found that ions are enhanced and neutrals are depleted in dense gas due to embedded CRs. Carbon chemistry is substantially altered depending on the assumed CR model: CRs produced by embedded sources create a significant reservoir of atomic carbon, mostly neutral, in dense gas. Embedded CRs reduce the amount of CO in clouds and warm the gas to over 30 K. In this letter we investigate the impact of the above effects on \XCI and \XCO. In Section \ref{sec:methods} we describe the methods used in this paper. In Section \ref{sec:res} we present the results and discuss the implications for observations.

\section{Methods}\label{sec:methods}
We use the same astrochemical models from \citetalias{gaches2019} and summarize the methodology here. See \citetalias{gaches2019} for further details.

We generate synthetic protoclusters assuming the Tapered Turbulent Core \citep{offner2011} accretion model following the method described in \citet{gaches2018a}. We directly sample from the bi-variate protostellar mass distribution using the method of conditional probabilities. Each molecular cloud is described by a gas surface density and number of constituent protostars, $\Sigma_{\rm cl}$ and $N_*$, respectively. We only consider models where the star formation efficiency, $\varepsilon_g \equiv M_*/M_{\rm gas} \leq 50\%$. 

We calculate the accelerated proton spectrum due to accretion shocks for each star in the protocluster. CR protons are assumed to be accelerated via Diffuse Shock Acceleration (DSA) \citep[reviewed by][]{drury1983, melrose2009} near the surface of the protostar. DSA predicts a power law spectra in momentum space, $j(p)$, with an injection momentum, $p_{\rm inj}$, set by the shock gas temperature and a maximum energy constrained by collisional energy losses and upstream diffusion \citep{gaches2018b}. The CR flux spectrum is
\begin{equation}
    j(p) = j_0 \left ( \frac{p}{p_{\rm inj}} \right )^{-a},
\end{equation}
where $j_0$ is the normalization constant calculated from the total shock energy and efficiency, and $a$ is set by the shock compression factor. We find that the maximum proton energy is typically between 1 - 10 GeV \citep{gaches2018b}. We attenuate the CRs by the gas surface density out of each protostellar core, $\Sigma_{\rm core} = 1.22 \Sigma_{\rm cl}$, following \citet{padovani2009}. We assume the CRs within the core free-stream outwards since shallower attenuation produces too much CR heating in the core (see \citealt{gaches2018b}). CRs may also be attenuated by the accretion flow (Offner et al. 2019, sub.), although we do not include this in the model. The total number of CRs escaping into a natal molecular cloud embedding a protocluster is the sum of the CRs accelerated by the individual protostars and then attenuated into the surrounding gas: $j_{\rm cluster}(E) = \sum_i^{N_*} j_i(E)$.

We embed the protoclusters in the center of one-dimensional molecular clouds with a density profile, $n(r) = n_0 (R/r)^{2}$. We set the outer density to $n_0 = 100$ cm$^{-3}$, and the radius, $R$, is determined by the total column density, $\mu m_H N(R) = \Sigma_{\rm cl}$. We utilize a modified version of the photo-dissociation region astrochemistry code {\sc 3d-pdr} \citep{bisbas2012} described in \citetalias{gaches2019}\footnote{The code is public at \url{https://uclchem.github.io/3dpdr.html}}, which includes CR attenuation in-situ. The astrochemistry code uses CR spectra at the surfaces of the gas model as inputs, rather than a global CRIR. It is not known exactly how CRs transport through molecular clouds. Therefore, we consider two different transport regimes: diffusive (1/r) and rectilinear (1/r$^2$). We use the two external CR spectra from \citet{ivlev2015}: a model that extrapolates the Voyager 1 data, $\mathcal{L}$, and one that attempts to account for modulation from interstellar gas, $\mathcal{H}$. 

We  also consider the impact of FUV radiation and we irradiate the external surface of the molecular cloud with the normalized interstellar radiation field described in \citet{draine1978}. We model the chemistry with the gas-phase {\sc Umist12} network \citep{mcelroy2013}, which includes 215 species and $\sim3000$ reactions. The network does not include gas-grain reactions, freeze-out or any desorption processes. We do not include the grain-assisted recombination proposed in the reduced network presented by \citet{gong2017}. We explored the impact of grain-assisted recombination for C$^+$ and He$^+$ on our results and found no significant changes in the CO- to CI-to-H$_2$ conversion factors. The inclusion of grain chemistry will be investigated in future studies. We include a model following the canonical setup: a low ionization rate with no attenuation, denoted as LNA. Models denoted with H or L utilize the $\mathcal{H}$ and $\mathcal{L}$ external spectra described above. Models without embedded sources are denoted with NI, while those including sources in the diffusive or rectiliniear regimes are denoted DI or RI, respectively. We consider the six different CR models listed in Table \ref{tab:models}: LNA, LNI, LRI, LDI, HNI, and HDI.

{\sc 3d-pdr} calculates the CO line-integrated emissivity, $\epsilon$, for the J-ladder from J=0 to J=41 and the CI 307 $\mu$m and 609 $\mu$m emissivities assuming non-local thermodynamic equilibrium and using an escape probability method to account for the line opactiy. We calculate the line flux from the emissivity:
\begin{equation}
    I = \frac{1}{2\pi} \int_0^R \epsilon \, dz ~~~ {\rm (erg \, s^{-1} \, cm^{-2}\, sr^{-1})},
\end{equation}
with 
\begin{equation}
    W = \frac{1}{10^5}\frac{c^3}{2 k_b \nu^3}I ~~~ {\rm (K \, km \, s^{-1})},
\end{equation}
where $c$ is the speed of light, $k_b$ is Boltzmann's constant and $\nu$ is the line frequency. This definition of integrated flux assumes that the interstellar medium is entirely optically thin. We calculate the H$_2$ column density from the astrochemical models
\begin{equation}
    N({\rm H_2}) = \int_0^R x({\rm H_2}) \, n_H \, dz,
\end{equation}
where $x({\rm H_2})$ is the abundance of H$_2$ and $n_H$ is the gas density. Finally, we compute \XCO using Equation \ref{eq:xco}.

\begin{deluxetable}{c|cccc}
\tablecolumns{5}
\tablecaption{Models from \citetalias{gaches2019}\label{tab:models}. L/H denotes a Low/High external CR spectrum, NI denotes no internal sources of CRs, DI denotes internal sources with $a = 1$ (diffusive transport), RI denotes internal sources with $a = 2$ (rectilinear transport) and NA denotes no internal sources or CR attenuation.}
\tablehead{\colhead{Name} & \colhead{Source Transport} & \colhead{Internal} & \colhead{External Field} & \colhead{Attenuation} }
\startdata
LDI\label{model:fid} & $r^{-1}$ & $\checkmark$ & $\mathcal{L}$ & $\checkmark$ \\
LRI\label{model:rec} & $r^{-2}$ & $\checkmark$ & $\mathcal{L}$ & $\checkmark$ \\
LNI\label{model:ni} & ... & ...  & $\mathcal{L}$ & $\checkmark$ \\
LNA\label{model:na} & ... & ...  & $\mathcal{L}$ & ...  \\
HNI\label{model:hni} & ... & ... & $\mathcal{H}$ & $\checkmark$ \\
HDI\label{model:hdi} & $r^{-1}$ & $\checkmark$ & $\mathcal{H}$ & $\checkmark$ \\
\enddata
\end{deluxetable}

\section{Results and Discussion}\label{sec:res}

We present the results from the astrochemical models on the CO-to-H$_2$  and CI-to-H$_2$  conversion factors here. A more general discussion on the astrochemical impact of CRs accelerated within protoclusters is presented in \citetalias{gaches2019}. 

\subsection{Effect of Cosmic Rays on \XCO}
\begin{figure*}[ht!]
    \centering
    \includegraphics[width=\textwidth]{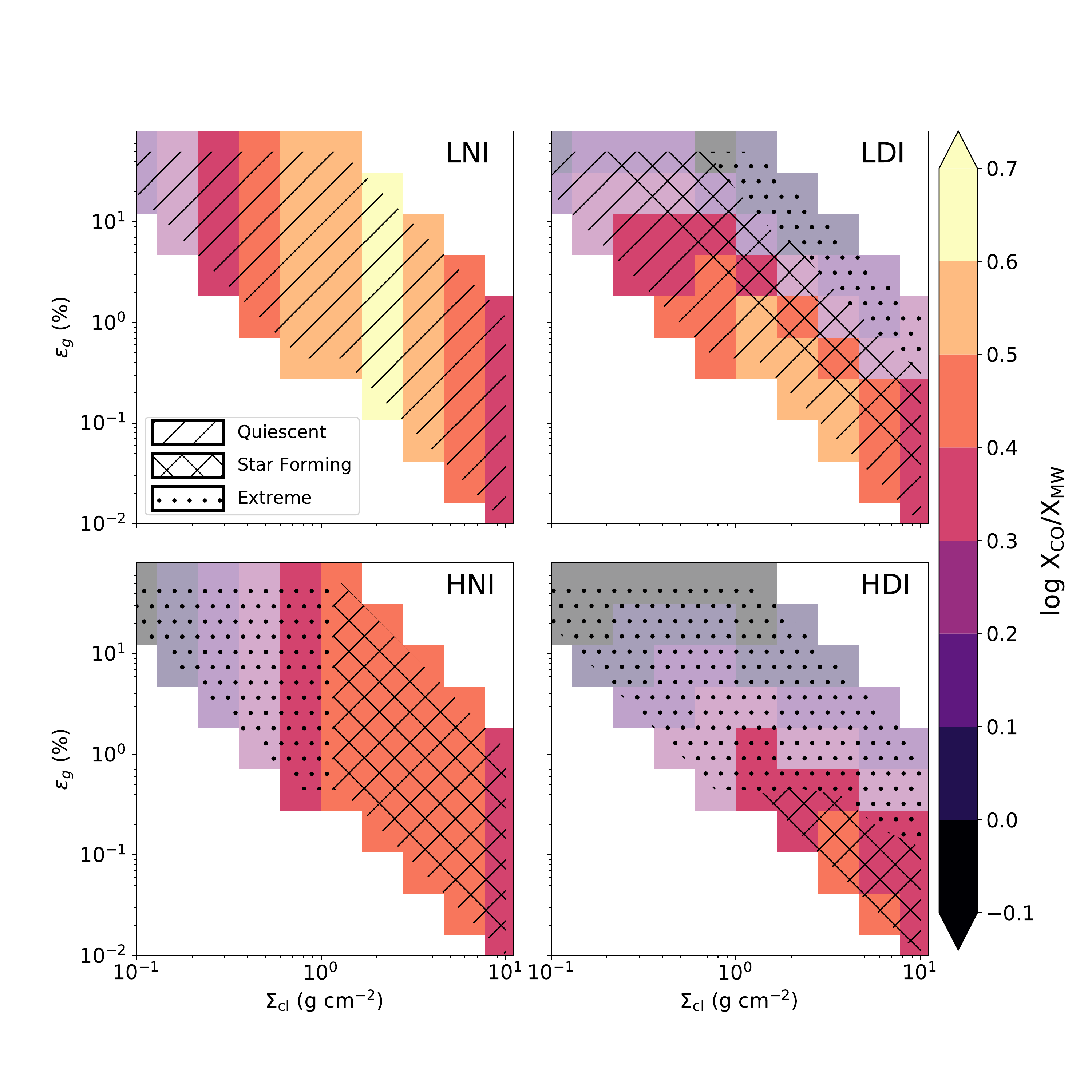}
    \caption{\label{fig:fideff} 
    Color shows $\log X_{\rm CO}/X_{\rm MW}$ where $X_{\rm MW} = 2 \times 10^{20}$ cm$^{-2}$ (K km s$^{-1}$)$^{-1}$ as a function of gas surface density, $\Sigma_{\rm cl}$, and star formation efficiency, $\varepsilon_g$. White shaded cells show regions where X$_{\rm CO}$ is consistent with Milky Way observations, $-0.3 \geq \log \XCO/X_{\rm MW} \leq 0.3$. The hatched regions indicate different cosmic-ray environments, where we define $\langle \zeta \rangle_x$, the spatially-averaged CRIR,  $\langle \zeta \rangle_x < 10^{-16}$ as ``quiescent'', $10^{-16} < \langle \zeta \rangle_x < 10^{-15}$ as ``star forming'' and $\langle \zeta \rangle_x > 10^{-15}$ as ``extreme.''}
\end{figure*}

Figure \ref{fig:fideff} shows the CO-to-H$_2$  conversion factor as a function of cloud surface density, $\Sigma_{\rm cl}$, and star formation efficiency, $\varepsilon_g$, for four of the CR models in 
Table \ref{tab:models}. We plot \XCO normalized to the fiducial MW value $X_{\rm MW} = 2 \times 10^{20}\,{\rm cm}^{-2}\,({\rm K}\,{\rm km}\,{\rm s}^{-1})^{-1}$ \citep{bolatto2013}. 

The behavior of \XCO changes significantly with the assumed CR model. \XCO varies only as function of surface density for the models without internal sources, LNI and HNI. There is a 0.2 dex offset in \XCO between models using the high and low external cosmic ray spectrum for $\Sigma_{\rm cl} < 3$ g cm$^{-2}$ owing to increased temperatures at low extinction in model HDI. The decline in \XCO at higher surface densities is the result of a larger turbulent line width because of two cooperating effects. First, there is a higher temperature due to the increasing importance of turbulent heating. Second, the turbulent linewidth produces brighter, but still optically thick, CO emission. 

In the models with CRs that attenuate diffusively, LDI and HDI, \XCO becomes a sensitive function of the star formation efficiency, losing much of the dependence on surface density. \XCO is reduced by up to 0.5 dex due to embedded sources with the lowest values occurring for the highest star formation efficiencies. It is important to emphasize that CRs from embedded sources do little to reduce the overall amount of H$_2$ \citepalias{gaches2019}. However, they cause two effects which act to decrease \XCO. First, while they reduce the amount of CO in deeply embedded regions of the cloud, they cause an enhancement of CO in low extinction gas due to an increase of HCO$^+$ and, following the formation of OH through H$_3^+$, the OH formation pathway becoming important \citep{bisbas2017}. Second, the increased CRIR leads to higher kinetic temperatures making the CO emission brighter overall. 

Some prior work has investigated the effect of star formation on \XCO. CR and FUV feedback from star formation external to the molecular cloud can be modeled by scaling their intensity linearly with the star formation rate (SFR) \citep{papadopoulos2010}. This is motivated by the relationship between the supernova rate and the SFR and implicitly assumes that CRs are mainly accelerated in supernova shocks. \citet{clark2015} used these relations to model how the SFR affects \XCO in simulated molecular clouds. They found that \XCO increases with the SFR if the cloud properties remain fixed. The increase of \XCO with SFR is very weak if the density of the cloud scales with the SFR. \citet{bisbas2015} modelled the effect of enhanced CRs on the ${\rm [CO/H_2]}$ ratio, comparing different environments. They show that ${\rm [CO/H_2]}$ decreases substantially with an increase in the CRIR. By construction, these models only account for variations in the external CR flux and neglect CRs accelerated within protoclusters due to accretion, jets or stellar winds. 

\subsection{Effect of Cosmic Rays on \XCI}
\begin{figure*}[ht!]
    \centering
    \includegraphics[width=\textwidth]{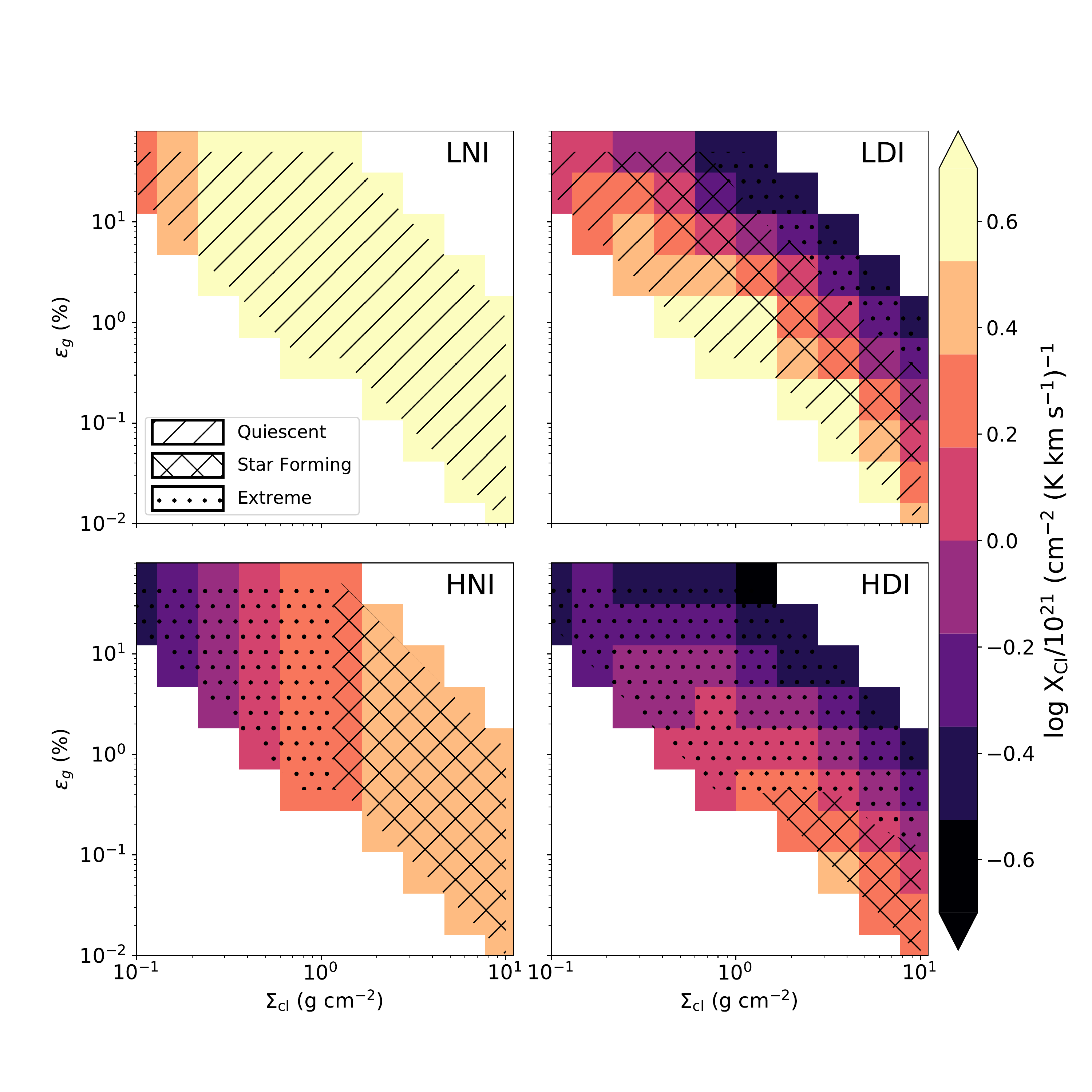}
    \caption{\label{fig:fidCIeff} Same as Figure \ref{fig:fideff} but for $\log X_{\rm CI}/10^{-21}$.}
\end{figure*}

Forbidden line emission from neutral carbon is a possible tracer for molecular gas, as discussed above. Figure \ref{fig:fidCIeff} shows \XCI as a function of surface density, $\Sigma_{\rm cl}$, and star formation efficiency, $\varepsilon_g$. \XCI shows the same qualitative trends as \XCO, although it is more sensitive to the CRIR: a spread of 1.2 dex in \XCI and 0.5 dex in \XCO for the LDI model. The canonical model, LNA, which has no attenuation, exhibits a maximum value of \XCI$ \geq 4 \times 10^{21}$ cm$^{-2}$ (K km s$^{-1}$)$^{-1}$. Models using the high, external CR spectrum, HNI and HDI, exhibit a 0.2-0.8 dex reduction in \XCI compared to the low spectrum models, LNI and LDI, respectively.

The increased CRIR throughout the cloud in the high models and those with internal sources causes atomic carbon to exist outside a thin transition layer. Atomic carbon is formed in the dense gas through the destruction of CO by He$^+$:
\begin{equation*}
    {\rm He + CR \rightarrow He^+ + e^-}
\end{equation*}
\begin{equation*}
    {\rm He^+ + CO \rightarrow He + O + C^+} 
\end{equation*}
with neutral carbon forming from recombination of C$^+$. Neutral carbon is also the result of direct dissociation of neutral molecules, such as CO, by CR protons and CR-generated photons. This enhancement leads to a reduced \XCI. Embedded sources cause \XCI to decrease by over an order of magnitude across two orders of magnitude increase in the star formation efficiency. 

Neutral carbon emission is easily observable at high redshifts due to the line shifting to millimeter wavelengths. Starburst galaxies have higher SFRs producing extreme environments and more CO-dark gas \citep{wolfire2010, glover2016}. Thus, at high redshifts and in galaxies undergoing starbursts, CI may become an optimal tracer of molecular gas.

\subsection{Statistical Trends} 
\begin{figure}
    \centering
    \includegraphics[trim=12 0 0 0, clip, width=\textwidth]{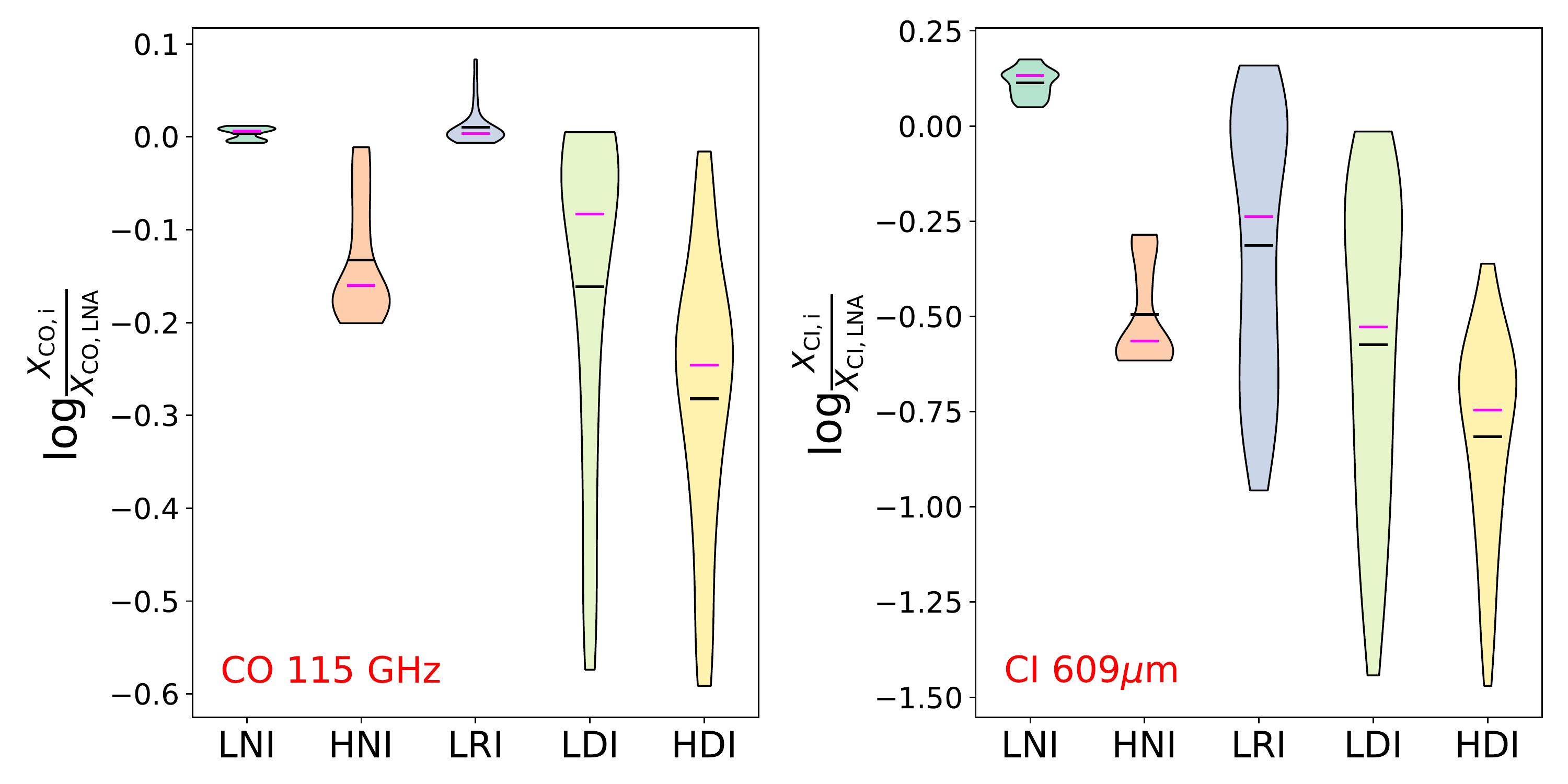}
    \caption{Logarithm of the ratio of X$_{\rm CO}$ and X$_{\rm CI}$ for a given cosmic-ray model $i \subset $ (LNI, HNI, LRI, LDI, HDI) compared to model LNA (low, external spectrum and no attenuation). Black line indicates the mean. Magenta line indicates the median.}
    \label{fig:violins}
\end{figure}

Figure \ref{fig:violins} statistically summarizes the impact of the various CR models on \XCI and \XCO. The violin plots show the distribution of the logarithmic difference between X$_{\rm i}$ as calculated with the canonical model, LNA, and each of the CR models in Table \ref{tab:models} using the clouds across the $\Sigma_{\rm cl}-\varepsilon_g$ space as samples. These distributions represent the impact on \XCO when CR attenuation or embedded sources are neglected. We find very little deviation in \XCO when attenuation is included in quiescent models without internal sources. Comparison to the star-forming and extreme CR model without internal sources, HNI, shows that \XCO will be over-estimated by 0.15 dex in calculations using the often-assumed CRIR of $\zeta \approx 10^{-17}$ s$^{-1}$. CRs from embedded sources, which propagate via diffusion, decrease \XCO for all clouds. Furthermore, there is a substantial spread due to variation with the number of protostars, $N_*$. The high model with internal sources, HDI, logarithmic difference with the canonical model exhibits a dispersion of 0.3 dex, similar to the spread derived from MW observations \citep{bolatto2013}. If CRs from embedded sources transport as r$^{-2}$ there is no impact on \XCO because the CRIR is lower and dominated by the CRs originating from external sources rather than internal.

The \XCI distributions in the right panel of Figure \ref{fig:violins} show much greater sensitivity to the CR model assumptions. All models differ significantly from the often-assumed canonical model in \XCI. \XCI decreases by 0.5 dex for the high model with no internal sources, HNI, and massive and inefficient star forming regions. In the case of a ``Quiescient'' CR environment,  CRs from embedded sources have a larger impact on \XCI. The inclusion of CRs from embedded sources in star-forming and extreme environments, represented by HDI, reduces \XCI by nearly a dex compared to the canonical model.

\subsection{Comparisons to Galactic-scale Observations}
The hatching in Figure \ref{fig:fideff} denotes different CR environments: ``Quiescent'' regions with $\langle \zeta \rangle_x < 10^{-16}$ s$^{-1}$, ``Star Forming'' regions with $10^{-16} < \langle \zeta \rangle_x < 10^{-15}$ s$^{-1}$ and ``Extreme'' regions with $\langle \zeta \rangle_x > 10^{-15}$ s$^{-1}$, where $\langle \zeta \rangle_x$ is the spatially-averaged CRIR. These labels are motivated by observational surveys which show the majority of pointings through diffuse gas have $10^{-16} < \zeta < 10^{-15}$ s$^{-1}$. Low A$_V$ observations where $\zeta > 10^{-15}$ s$^{-1}$ are primarily sight-lines towards the galactic center \citep{indriolo2012, indriolo2015}.

There have been numerous observational studies measuring \XCO in different environments within the MW and other galaxies \citep[see][and citations within]{bolatto2013}. Remarkably, in the MW and many of the Local Group galaxies, \XCO is relatively constant on kpc scales. The consistency of \XCO in the MW and Local Group can be explained by similar molecular cloud properties due to star-formation feedback \citep{narayanan2013}. There is a general trend in star-forming galaxies of low values of \XCO towards the center and larger values in the outer disk \citep{sandstrom2013}. 

The white shading in Figure \ref{fig:fideff} shows where \XCO is consistent with the MW average value and spread. Models without embedded sources, LNI and HNI, are only consistent with the MW value for $\Sigma_{\rm cl} < 0.2$ g cm$^{-2}$ and $\Sigma_{\rm cl} < 0.6$ g cm$^{-2}$, respectively. Models with high surface density and low star formation efficiency, similar to clouds in the Galactic Center, exhibit a decreased \XCO compared to clouds with $\Sigma \approx 1$ g cm$^{-2}$. The introduction of embedded sources increases the agreement with the MW \XCO. Clouds with star formation efficiencies greater than 2\% in the low model with internal sources, LDI, have $\XCO = (2.5 \pm 1)\times10^{20}$ cm$^{-2}$(K km s$^{-1}$)$^{-1}$, consistent with the MW value. High models with the internal sources show $\XCO = 3 \pm 1.5 \times 10^{20}$ cm$^{-2}$(K km s$^{-1}$)$^{-1}$ .{\it Thus, CRs accelerated during the star formation process act to regulate \XCO and reduce variation}.

Starburst galaxies tend to have lower values of \XCO \citep{downes1998, papadopoulos1999, papadopoulos2012, salak2014}. Our models show that \XCO always decreases towards regions with more extreme CR environments. Environmental changes, which occur in higher redshift galaxies due to enhanced supernova rates, will also decrease \XCO and \XCI. In starburst galaxies, which have high star formation rates, this decrease could be compounded by CRs produced during the star formation process.

\subsection{Summary}
We found in \citetalias{gaches2019} that the inclusion of CR sources, specifically accreting protostars, embedded within molecular clouds and CR attenuation make the CRIR vary throughout the cloud. In this paper, we investigate the impact of different external CR fluxes and the inclusion of embedded CR sources on the CO-to-H$_2$ and CI-to-H$_2$ conversion factors. We find that differences in the CR flux caused by changes in the external environment and embedded star formation alter \XCI significantly and \XCO by factors of a few. However, external environmental changes alone reduce \XCO only by 0.2 dex, within the measured spread of \XCO in the MW \citep{bolatto2013}. The difference in \XCI is more pronounced:  it declines by an order of magnitude for the lowest surface density environments. The inclusion of embedded CR sources removes the strong dependence of \XCO and \XCI on surface density and reduces the conversion factors by 0.6 and 1.2 dex, respectively. Embedded sources act to regulate \XCO and reduce variation as a function of gas surface density and star formation efficiency. Clouds in low model including embedded sources, LDI, with star formation efficiences greater than 2\% are consistent with the observed MW value and spread of $X_{\rm CO, MW} = 2 \times 10^{20} \pm 0.3$ dex cm$^{-2}$ (K km s$^{-1}$)$^{-1}$. Observations of the CRIR in diffuse gas in the MW show that the average CRIR, $\langle \zeta \rangle \approx 10^{-16}$, which is represented by our models with a high surface CR spectrum. Models with this CRIR and ongoing star formation, are consistent with the observed MW value for all regions with star formation efficiencies greater than 1\%. Our models reproduce the trends of a decreasing \XCO towards more extreme CR environments, such as those observed in the Galactic Center, the high redshift universe and starburst galaxies. Our results motivate the inclusion of CR physics and the possibility of cosmic-ray feedback from internal sources when modeling \XCO and \XCI.

\software{ \begin{itemize} \item {\sc 3d-pdr} \citep{bisbas2012}, CR implementation \citepalias{gaches2019} 
\item {\sc matplotlib} \citep{matplotlib2007} 
\item {\sc NumPy} \citep{numpy2011} 
\item {\sc SciPy} \citep{scipy2001} 
\item {\sc JupyterLab} \end{itemize}}

\acknowledgments
SSRO acknowledges funding by the National Science Foundation (NSF) grants: NSF AAG grant AST-1510021 and NSF CAREER grant AST-1650486. TGB acknowledges funding by the German Science Foundation (DFG) via the Collaborative Research Center SFB 956 ``Conditions and impact of star formation''. The authors thank the anonymous referee for their useful comments. The calculations performed for this work were done on the Massachusetts Green High Performance Computing Center (MGHPCC) in Holyoke, Massachusetts supported by the University of Massachusetts, Boston University, Harvard University, MIT, Northeastern University and the Commonwealth of Massachusetts.

\iffalse
\appendix
\section{Extra Plots}
\begin{figure}
    \centering
    \includegraphics[width=\textwidth]{WCO_violins.pdf}
    \caption{Logarithm of the ratio of W$_{\rm CO}$, N$_{\rm CO}$, and $\left < T_x \right >$ on the left, middle and right, respectively, for a given cosmic-ray model $i \subset $ (LNI, HNI, LRI, LDI, HDI) with model LNA. Black line indicates the mean. Magenta line indicates the median.}
    \label{fig:violins_wco}
\end{figure}
\fi

\bibliography{cits}

\begin{thebibliography}{}
\expandafter\ifx\csname natexlab\endcsname\relax\def\natexlab#1{#1}\fi

\bibitem[{{Bell} {et~al.}(2006){Bell}, {Roueff}, {Viti}, \&
  {Williams}}]{bell2006}
{Bell}, T.~A., {Roueff}, E., {Viti}, S., \& {Williams}, D.~A. 2006, \mnras,
  371, 1865

\bibitem[{{Bisbas} {et~al.}(2012){Bisbas}, {Bell}, {Viti}, {Yates}, \&
  {Barlow}}]{bisbas2012}
{Bisbas}, T.~G., {Bell}, T.~A., {Viti}, S., {Yates}, J., \& {Barlow}, M.~J.
  2012, \mnras, 427, 2100

\bibitem[{{Bisbas} {et~al.}(2015){Bisbas}, {Papadopoulos}, \&
  {Viti}}]{bisbas2015}
{Bisbas}, T.~G., {Papadopoulos}, P.~P., \& {Viti}, S. 2015, \apj, 803, 37

\bibitem[{{Bisbas} {et~al.}(2017){Bisbas}, {van Dishoeck}, {Papadopoulos},
  {Sz{\H u}cs}, {Bialy}, \& {Zhang}}]{bisbas2017}
{Bisbas}, T.~G., {van Dishoeck}, E.~F., {Papadopoulos}, P.~P., {et~al.} 2017,
  \apj, 839, 90

\bibitem[{{Bolatto} {et~al.}(2013){Bolatto}, {Wolfire}, \&
  {Leroy}}]{bolatto2013}
{Bolatto}, A.~D., {Wolfire}, M., \& {Leroy}, A.~K. 2013, \araa, 51, 207

\bibitem[{{Clark} \& {Glover}(2015)}]{clark2015}
{Clark}, P.~C., \& {Glover}, S. C.~O. 2015, \mnras, 452, 2057

\bibitem[{{Downes} \& {Solomon}(1998)}]{downes1998}
{Downes}, D., \& {Solomon}, P.~M. 1998, \apj, 507, 615

\bibitem[{{Draine}(1978)}]{draine1978}
{Draine}, B.~T. 1978, \apjs, 36, 595

\bibitem[{{Drury}(1983)}]{drury1983}
{Drury}, L.~O. 1983, Reports on Progress in Physics, 46, 973

\bibitem[{{Gaches} \& {Offner}(2018{\natexlab{a}})}]{gaches2018a}
{Gaches}, B.~A.~L., \& {Offner}, S.~S.~R. 2018{\natexlab{a}}, \apj, 854, 156

\bibitem[{{Gaches} \& {Offner}(2018{\natexlab{b}})}]{gaches2018b}
---. 2018{\natexlab{b}}, \apj, 861, 87

\bibitem[{{Gaches} {et~al.}(2019){Gaches}, {Offner}, \& {Bisbas}}]{gaches2019}
{Gaches}, B. A.~L., {Offner}, S. S.~R., \& {Bisbas}, T.~G. 2019, arXiv
  e-prints, arXiv:1905.02232

\bibitem[{{Glover} \& {Clark}(2016)}]{glover2016}
{Glover}, S. C.~O., \& {Clark}, P.~C. 2016, \mnras, 456, 3596

\bibitem[{{Glover} {et~al.}(2015){Glover}, {Clark}, {Micic}, \&
  {Molina}}]{glover2015}
{Glover}, S. C.~O., {Clark}, P.~C., {Micic}, M., \& {Molina}, F. 2015, \mnras,
  448, 1607

\bibitem[{{Gong} {et~al.}(2018){Gong}, {Ostriker}, \& {Kim}}]{gong2018}
{Gong}, M., {Ostriker}, E.~C., \& {Kim}, C.-G. 2018, \apj, 858, 16

\bibitem[{{Gong} {et~al.}(2017){Gong}, {Ostriker}, \& {Wolfire}}]{gong2017}
{Gong}, M., {Ostriker}, E.~C., \& {Wolfire}, M.~G. 2017, \apj, 843, 38

\bibitem[{{Hollenbach} \& {Tielens}(1999)}]{hollenbach1999}
{Hollenbach}, D.~J., \& {Tielens}, A.~G.~G.~M. 1999, Reviews of Modern Physics,
  71, 173

\bibitem[{{Hunter}(2007)}]{matplotlib2007}
{Hunter}, J.~D. 2007, Computing in Science and Engineering, 9, 90

\bibitem[{{Ikeda} {et~al.}(1999){Ikeda}, {Maezawa}, {Ito}, {Saito}, {Sekimoto},
  {Yamamoto}, {Tatematsu}, {Arikawa}, {Aso}, {Noguchi}, {Shi}, {Miyazawa},
  {Saito}, {Ozeki}, {Fujiwara}, {Ohishi}, \& {Inatani}}]{ikeda1999}
{Ikeda}, M., {Maezawa}, H., {Ito}, T., {et~al.} 1999, \apj, 527, L59

\bibitem[{{Indriolo} \& {McCall}(2012)}]{indriolo2012}
{Indriolo}, N., \& {McCall}, B.~J. 2012, \apj, 745, 91

\bibitem[{{Indriolo} {et~al.}(2015){Indriolo}, {Neufeld}, {Gerin}, {Schilke},
  {Benz}, {Winkel}, {Menten}, {Chambers}, {Black}, {Bruderer}, {Falgarone},
  {Godard}, {Goicoechea}, {Gupta}, {Lis}, {Ossenkopf}, {Persson},
  {Sonnentrucker}, {van der Tak}, {van Dishoeck}, {Wolfire}, \&
  {Wyrowski}}]{indriolo2015}
{Indriolo}, N., {Neufeld}, D.~A., {Gerin}, M., {et~al.} 2015, \apj, 800, 40

\bibitem[{{Ivlev} {et~al.}(2015){Ivlev}, {Padovani}, {Galli}, \&
  {Caselli}}]{ivlev2015}
{Ivlev}, A.~V., {Padovani}, M., {Galli}, D., \& {Caselli}, P. 2015, \apj, 812,
  135

\bibitem[{Jones {et~al.}(2001--)Jones, Oliphant, Peterson,
  {et~al.}}]{scipy2001}
Jones, E., Oliphant, T., Peterson, P., {et~al.} 2001--, {SciPy}: Open source
  scientific tools for {Python}, , , [Online;]

\bibitem[{{Kennicutt} \& {Evans}(2012)}]{kennicutt2012}
{Kennicutt}, R.~C., \& {Evans}, N.~J. 2012, \araa, 50, 531

\bibitem[{{Kulesa} {et~al.}(2005){Kulesa}, {Hungerford}, {Walker}, {Zhang}, \&
  {Lane}}]{kulesa2005}
{Kulesa}, C.~A., {Hungerford}, A.~L., {Walker}, C.~K., {Zhang}, X., \& {Lane},
  A.~P. 2005, \apj, 625, 194

\bibitem[{{Lagos} {et~al.}(2012){Lagos}, {Bayet}, {Baugh}, {Lacey}, {Bell},
  {Fanidakis}, \& {Geach}}]{lagos2012}
{Lagos}, C.~d.~P., {Bayet}, E., {Baugh}, C.~M., {et~al.} 2012, \mnras, 426,
  2142

\bibitem[{{Lo} {et~al.}(2014){Lo}, {Cunningham}, {Jones}, {Bronfman}, {Cortes},
  {Simon}, {Lowe}, {Fissel}, \& {Novak}}]{lo2014}
{Lo}, N., {Cunningham}, M.~R., {Jones}, P.~A., {et~al.} 2014, \apj, 797, L17

\bibitem[{{McElroy} {et~al.}(2013){McElroy}, {Walsh}, {Markwick}, {Cordiner},
  {Smith}, \& {Millar}}]{mcelroy2013}
{McElroy}, D., {Walsh}, C., {Markwick}, A.~J., {et~al.} 2013, \aap, 550, A36

\bibitem[{Melrose(2009)}]{melrose2009}
Melrose, D.~B. 2009, in Encyclopedia of Complexity and Systems Science, ed.
  R.~A. Meyers (Springer), 21--42

\bibitem[{{Narayanan} \& {Hopkins}(2013)}]{narayanan2013}
{Narayanan}, D., \& {Hopkins}, P.~F. 2013, \mnras, 433, 1223

\bibitem[{{Offner} {et~al.}(2014){Offner}, {Bisbas}, {Bell}, \&
  {Viti}}]{offner2014}
{Offner}, S.~S.~R., {Bisbas}, T.~G., {Bell}, T.~A., \& {Viti}, S. 2014, \mnras,
  440, L81

\bibitem[{{Offner} \& {McKee}(2011)}]{offner2011}
{Offner}, S. S.~R., \& {McKee}, C.~F. 2011, \apj, 736, 53

\bibitem[{{Padovani} {et~al.}(2009){Padovani}, {Galli}, \&
  {Glassgold}}]{padovani2009}
{Padovani}, M., {Galli}, D., \& {Glassgold}, A.~E. 2009, \aap, 501, 619

\bibitem[{{Papadopoulos}(2010)}]{papadopoulos2010}
{Papadopoulos}, P.~P. 2010, \apj, 720, 226

\bibitem[{{Papadopoulos} {et~al.}(2018){Papadopoulos}, {Bisbas}, \&
  {Zhang}}]{papadopoulos2018}
{Papadopoulos}, P.~P., {Bisbas}, T.~G., \& {Zhang}, Z.-Y. 2018, \mnras, 478,
  1716

\bibitem[{{Papadopoulos} \& {Seaquist}(1999)}]{papadopoulos1999}
{Papadopoulos}, P.~P., \& {Seaquist}, E.~R. 1999, \apj, 516, 114

\bibitem[{{Papadopoulos} {et~al.}(2004){Papadopoulos}, {Thi}, \&
  {Viti}}]{papadopoulos2004}
{Papadopoulos}, P.~P., {Thi}, W.~F., \& {Viti}, S. 2004, \mnras, 351, 147

\bibitem[{{Papadopoulos} {et~al.}(2012){Papadopoulos}, {van der Werf},
  {Xilouris}, {Isaak}, \& {Gao}}]{papadopoulos2012}
{Papadopoulos}, P.~P., {van der Werf}, P., {Xilouris}, E., {Isaak}, K.~G., \&
  {Gao}, Y. 2012, \apj, 751, 10

\bibitem[{{Remy} {et~al.}(2017){Remy}, {Grenier}, {Marshall}, \&
  {Casandjian}}]{remy2017}
{Remy}, Q., {Grenier}, I.~A., {Marshall}, D.~J., \& {Casandjian}, J.~M. 2017,
  \aap, 601, A78

\bibitem[{{Riechers} {et~al.}(2019){Riechers}, {Pavesi}, {Sharon}, {Hodge},
  {Decarli}, {Walter}, {Carilli}, {Aravena}, {da Cunha}, {Daddi}, {Dickinson},
  {Smail}, {Capak}, {Ivison}, {Sargent}, {Scoville}, \& {Wagg}}]{riechers2019}
{Riechers}, D.~A., {Pavesi}, R., {Sharon}, C.~E., {et~al.} 2019, \apj, 872, 7

\bibitem[{{Salak} {et~al.}(2014){Salak}, {Nakai}, \& {Kitamoto}}]{salak2014}
{Salak}, D., {Nakai}, N., \& {Kitamoto}, S. 2014, \pasj, 66, 96

\bibitem[{{Sandstrom} {et~al.}(2013){Sandstrom}, {Leroy}, {Walter}, {Bolatto},
  {Croxall}, {Draine}, {Wilson}, {Wolfire}, {Calzetti}, {Kennicutt}, {Aniano},
  {Donovan Meyer}, {Usero}, {Bigiel}, {Brinks}, {de Blok}, {Crocker}, {Dale},
  {Engelbracht}, {Galametz}, {Groves}, {Hunt}, {Koda}, {Kreckel}, {Linz},
  {Meidt}, {Pellegrini}, {Rix}, {Roussel}, {Schinnerer}, {Schruba}, {Schuster},
  {Skibba}, {van der Laan}, {Appleton}, {Armus}, {Brandl}, {Gordon}, {Hinz},
  {Krause}, {Montiel}, {Sauvage}, {Schmiedeke}, {Smith}, \&
  {Vigroux}}]{sandstrom2013}
{Sandstrom}, K.~M., {Leroy}, A.~K., {Walter}, F., {et~al.} 2013, \apj, 777, 5

\bibitem[{{Shetty} {et~al.}(2011){Shetty}, {Glover}, {Dullemond}, \&
  {Klessen}}]{shetty2011}
{Shetty}, R., {Glover}, S.~C., {Dullemond}, C.~P., \& {Klessen}, R.~S. 2011,
  \mnras, 412, 1686

\bibitem[{{van der Walt} {et~al.}(2011){van der Walt}, {Colbert}, \&
  {Varoquaux}}]{numpy2011}
{van der Walt}, S., {Colbert}, S.~C., \& {Varoquaux}, G. 2011, Computing in
  Science and Engineering, 13, 22

\bibitem[{{Wolfire} {et~al.}(2010){Wolfire}, {Hollenbach}, \&
  {McKee}}]{wolfire2010}
{Wolfire}, M.~G., {Hollenbach}, D., \& {McKee}, C.~F. 2010, \apj, 716, 1191

\end{thebibliography}

\end{document}